\documentclass[final,5p,times,twocolumn]{elsarticle}

\usepackage{graphicx}

\usepackage{amssymb}

\biboptions{}

\journal{Solid State Communications}

\begin{document}

\begin{frontmatter}


\title{Title\tnoteref{label1}}
\author{Rayda Gammag\corref{cor1}}
\ead{rayda.gammag@up.edu.ph}
\author{Cristine Villagonzalo}
\cortext[cor1]{Corresponding author}
\address{Structure and Dynamics Group, National Institute of Physics\\ University of the Philippines, 
Diliman, 1101 Quezon City, Philippines}

\title{Quenching of the DOS beats in a two-dimensional electron gas in tilted magnetic fields}




\begin{abstract}
A two-dimensional electron gas exposed to a tilted magnetic field is considered with the Rashba spin-orbit 
interaction and the Zeeman effect.  An exact solution for the eigenvalues was obtained assuming that 
two opposite spin states of adjacent Landau levels have equal probability.  No crossings between adjacent 
eigenenergies were observed, for the tilt angles studied here ($\theta \leq 80^{\circ}$), 
unlike in the perpendicular-magnetic-field case. The absence of crossings lead 
to quenched beating structures in the oscillations of the density of states (DOS). Persistent spin-splittings 
were observed at the weak magnetic field region.  The splittings, however, can be effectively screened by an 
increased Landau level broadening.  The results shed light on how spins can be controlled through the Rashba 
interaction strength, the disorder-related broadening and the magnetic field tilt angle.
\end{abstract}


\end{frontmatter}


\section{Introduction}
The focus and interest in the development of spintronics lies in the maneuvering of the spin, rather than solely the charge, 
as the medium of transport in microelectronics \cite{Grundler2002, Datta1989, Nitta2007}. For instance, the 
spin field-effect transistor (spin-FET) proposed by Datta and Das consists of a two-dimensional electron gas (2DEG) 
connected to ferromagnetic electrodes \cite{Datta1989}.  Here spin-polarized electrons from a ferromagnetic source 
electrode are injected into the 2DEG.  The spin can be rotated because an effective magnetic field, modulated by 
the gate voltage, is induced by the spin-orbit interaction.  

Rashba first studied a wurtzite lattice semiconductor where an energy extremum is reached not at isolated points 
in the Brillouin zone but over an extended region \cite{Rashba1960}.  He considered that spin-orbit interaction 
(SOI) serves as a measure of the depth of this extremum loop.  Using the effective mass approximation, Rashba 
obtained the eigenvalues of electrons with SOI under a perpendicular magnetic field.  His work has since then 
served as a benchmark of SOI studies arising from inversion asymmetric potentials.

Whenever a 2DEG layer is sandwiched in a host system that lacks structure inversion symmetry the spin degeneracy of 
the electron states is lifted even in the absence of an external magnetic field \cite{Luo1990}.  This zero-magnetic-field 
spin splitting is relativistic in nature.  With respect to a laboratory frame, the electrons are moving in an electric 
field brought about by the potential well asymmetry.  In the electrons' rest frame, however, it is the electric field 
that is varying.  Hence, the electric field appears to the electrons as a magnetic field, which 
is now called the Rashba field $B_R$.  This nonzero $B_R$ has been measured, for example, in GaAs and InSb and other 
III-V heterostructures and in some narrow-gap II-VI semiconductor devices \cite{Fujii2002, Giglberger2007, Meier2007, Gui2004E}.

Most research investigations on 2DEG with Rashba SOI have dealt with the case when an external magnetic field 
$\vec{B}$ is oriented perpendicular to the plane \cite{WangVa2003,Cho2007}.  A common feature of these studies 
is the manifestation of beating patterns in the Shubnikov-de Haas (SdH) oscillations which is attributed to the 
unequally spaced energy spectrum brought about by the Rashba SOI \cite{Averkiev2005}.  In general, however, 
there are two types of spin-orbit interactions that govern the nature of a 2DEG found in heterostructures, 
namely, the Rashba and the Dresselhaus terms. The latter's electric field is associated with the host crystal's 
inversion asymmetry and is enhanced by the confinement \cite{Luo1990} and by temperature \cite{Gui2004}. 
The Rashba contribution is usually stronger as evidenced by III - V semiconductors \cite{Datta1989, Giglberger2007} 
and also in HgTe quantum wells \cite{Gui2004}.  When only one type of SOI is present or when one dominates the other, 
the beats in the oscillations are expected. But when the two SOI contributions are equal, the SdH oscillations 
are described by a singular frequency, that is, the beats are suppressed.  The absence of the beating pattern 
in the resistivity versus $B$, for example, is attributed to equidistant energy levels \cite{Averkiev2005}.

Although most studies dwelt on the perpendicular-magnetic-field problem \cite{WangVa2003,Wang2009,Shen2004,Shen2005}, 
some also explored the effects of tilting the orientation of $B$ with respect to the 2DEG plane.  In an 
experimental work on an AlGaAs/GaAs heterojunction, for example, it was found that the magnetization versus 
$B_z$ exhibits beating patterns at large tilt angles \cite{Wilde2009}.  On the contrary, a collapse of the 
ringlike structures in the longitudinal resistivity was observed for increasing tilt angles.  The ringlike 
structures originate from the crossings of two sets of spin split Landau levels from different sub-bands 
\cite{Ferreira2010}.  A theoretical work, on the other hand, found that when a 2DEG confined to a parabolic 
potential is under a tilted $\vec{B}$, its magnetoresistance exhibit beating patterns \cite{Zhang1997}. 
The beats occurred when the cyclotron frequency is less than the frequency of the confining potential 
and $\vec{B}$ is nearly parallel to the 2DEG plane.  Another research group observed a beating pattern 
in the Fermi energy versus the tilt angle of a 2DEG under a parabolic potential, free of spin-orbit 
interaction but inclusive of Zeeman splitting \cite{Ramos2009}.  When the same group included the Rashba 
SOI no pronounced beats were observed.  In between adjacent quantum Hall plateaus, they found that in 
the presence of Rashba SOI or tilted fields, new intermediate plateaus arise.  These are attributed to 
the lifted degeneracy \cite{Alves2009}.  

The aim of this work is to solve the eigenvalues of a 2DEG with Rashba and Zeeman interactions when the external 
magnetic field is applied at an angle relative to the gas plane's normal. No exact solution to this problem is 
established yet.  After obtaining the eigenenergies, the density of states will be calculated numerically in 
order to compare the results with experimental data and other theoretical studies. 

\section{Theoretical Formalism: An Analytic Solution}

\subsection{The Rashba SOI Hamiltonian}

The Hamiltonian of an electron in the presence of a magnetic field $\vec{B} = (B_x, B_y, B_z)$ can be written as
\begin{equation}
H = H_0 + H_R + H_Z = \frac{\hbar^2 \vec{k}^2}{2m^*} +  \alpha (\vec{\sigma} \times \vec{k}) \cdot \hat{z} - \vec{\mu} \cdot \vec{B},
\label{H}
\end{equation}
\noindent where $H_0$ is the free particle energy, $H_R$ is the Rashba spin-orbit interaction, and $H_Z$ is the 
Zeeman energy.  Here $m^*$ is the electron's effective mass. The strength of the Rashba SOI is indicated by 
the parameter $\alpha$ which is assumed to be constant for this present work. Here $\vec{\sigma}$ are the 
Pauli matrices and $\vec{k}$ is the wave vector.  The magnitude of the latter is determined by 
$k_j = -i \nabla_j + \frac{e}{ \hbar} A_j$, where $e$ is the electronic charge, $\hbar$ is Planck's 
constant $h$ over $2 \pi$, and $A_j$ is the $j$-th component of the magnetic vector potential.

One expects from Eq.~(\ref{H}) that for strong $\vec{B}$ the Zeeman term dominates while the Rashba term leads 
in the weak field region. However, in HgTe quantum wells, SOI is larger or comparable to the \textit{sp-d} 
exchange-interaction-induced giant Zeeman splitting \cite{Gui2004}. Note that the spin-splitting brought 
about by $H_R$ does not require an external $\vec{B}$ unlike $H_Z$.  The magnitude of $H_R$ is determined 
by the potential well asymmetry or by an external strain \cite{Rashba2006} while that of $H_Z$ is 
regulated mainly by the intensity of $\vec{B}$. 

Electrons in the $x-y$ plane, when subject to a perpendicular field of magnitude $B$, follow cyclotron 
orbits with angular frequency $\omega_c = eB/m^*$.
Hence, they are essentially under a harmonic oscillator potential.  Here we followed Rashba's formulation 
\cite{Rashba1960} where we let $k_{\pm} = k_x \pm i k_y$. When $\vec{B}$ is tilted at an angle $\theta$ 
with respect to the perpendicular $z$-axis, the commutation of the wave vectors depends on the magnitude 
and direction, that is,
\begin{equation}
[k_x, k_y] = - \frac{ieB_z}{\hbar} = - \frac{ieB}{\hbar} \, \mbox{cos} \, \theta.  
\end{equation}
\noindent When $\theta = 0$, we obtain Rashba's results that $[k_x, k_y]$ depends on $B$ only.  Moreover, 
since the momentum components do not commute they represent incompatible observables with an underlying 
uncertainty principle \cite{Shankar1994}. Relating the wave vector to the momentum by $p_i = \hbar k_i$, 
the uncertainty in simultaneous measurements of the momentum components becomes 
\begin{equation}
 \sigma_{p_x} \, \sigma_{p_y} \ge \frac{1}{2}\left(\frac{\hbar eB_z}{c}\right).
 \label{Eq:uncertainty}
\end{equation}

Keeping this in mind, the analytic solution can be conveniently obtained by using the ladder operators
\begin{equation}
a = \sqrt{\frac{\hbar c}{2 e B_z}} k_- \qquad \mbox{and} \qquad a^{\dagger} = \sqrt{\frac{\hbar c}{2 e B_z}} k_+, 
\label{Eq:ladders}
\end{equation}
\noindent which satisfy the canonical commutation relation $[a, a^{\dagger}] = 1$.  

The Hamiltonian can be rewritten as
\begin{equation}
H = \left( \begin{array}{cc}
H_n  + \Omega_z & i \Upsilon a + \Omega_-  \\ 
-i \Upsilon a^{\dagger} + \Omega_+   & H_n - \Omega_z           
\end{array}
  \right)
\end{equation}
\noindent where $H_n = \frac{1}{2}(a a^{\dagger} + a^{\dagger} a)$, $\Omega_j = B_jm^* / 2B_zm_e$ is the $j$-th 
component of the rescaled Zeeman energy, $\Upsilon = 2 \alpha \sqrt{\zeta / \hbar \omega_c}$ is the rescaled 
Rashba parameter, $\zeta = m^*/2 \hbar^2$ and $\Omega_{\pm} = \Omega_x \pm i \Omega_y$.  Here we have assumed 
that the g-factor $g = 2$.   In this work, energy units are given in terms of $\hbar \omega_c$ where $\omega_c = eB_z/m^*$.
We used $m^* = 0.05m_e$ where $m_e$ is the rest mass of an electron.

In solving the Schr\"odinger equation,
\begin{equation}
\left ( \begin{array}{cc}
H_n  + \Omega_z - E & i \Upsilon a + \Omega_-  \\ 
-i \Upsilon a^{\dagger} + \Omega_+   & H_n - \Omega_z - E  \end{array} 
\right) \left( \begin{array}{c} \sum_{n = 0} a_n \phi_n\\ 
\sum_{n = 0} b_n \phi_n  \end{array} \right) = 0,
\end{equation}
\noindent we express the wave function solutions in terms of a superposition of the harmonic oscillator basis 
functions $\phi_n$ where the coefficients $a_n$ and $b_n$ are the unknown spin-up and spin-down complex 
coefficients of the $n$th Landau level, respectively.  This yields a set of secular equations
\begin{equation}
\left (n + \frac{1}{2} + \Omega_z - E \right) a_n + i \Upsilon \sqrt{n + 1} \, b_{n + 1} + \Omega_- b_n = 0,
\label{Eq:secular1}
\end{equation}
\begin{equation}
 -i \Upsilon \sqrt{n} \, a_{n - 1} + \Omega_+ a_n + \left(n + \frac{1}{2} - \Omega_z - E \right) b_n = 0.
\label{Eq:secular2}
\end{equation}
\noindent Note that the secular equations derived by Rashba \cite{Rashba1960} are recovered when $\vec{B}$ 
is normal to the 2DEG plane, that is, when we let $\Omega_+ = \Omega_- = 0$.
We will impose this latter condition on secular Eqs.~(\ref{Eq:secular1}) and (\ref{Eq:secular2}) to obtain
the 2DEG behavior when $\theta = 0$. Later we will be needing such information 
for comparison with our results on tilted fields.

When we consider the pure Zeeman case ($\Upsilon = 0$), the eigenvalues are shown to be
\begin{equation}
 E_Z^{\pm} = n + \frac{1}{2} \pm \Omega.
\label{Eq:EZ}
\end{equation}
\noindent Accounting for the contribution of the in-plane magnetic field, the general Zeeman term 
$\Omega = \sqrt{\Omega_x^2 + \Omega_y^2 + \Omega_z^2}$ replaces the $\Omega_z$ in the 
perpendicular-magnetic-field case.  As expected, the spin-splitting brought about by the Zeeman 
effect $\Delta_Z = 2 \Omega$ is determined by the magnitude of $\vec{B}$ which includes both the 
in-plane and perpendicular components. On the other hand, when only the Rashba SOI is the source 
of spin splitting ($\Omega = 0$), the eigenvalues can be expressed as
\begin{equation}
 E_R^{\pm} = n \pm \frac{1}{2} \sqrt{1 + 4 \Upsilon^2 n}.
\end{equation}
\noindent The spin splitting brought about solely by the Rashba SOI 
$\Delta_R = E_R^+ - E_R^- = \sqrt{1 + 4 \Upsilon^2 n}$ is directly proportional to the strength 
of the spin-orbit interaction and to the square root of Landau level index.  The number of filled 
Landau levels $n$ is inversely proportional to $B_z = B$cos$\theta$. This means that for a given 
direction $\theta$, the Rashba splitting is more pronounced as $B \rightarrow 0$.  When the 
intensity $B$ is fixed, $\Delta_R$ becomes larger as $\theta$ approaches $90^{\circ}$.

\subsection{The Energy Spectrum}
We need to solve Eq.\ (\ref{Eq:secular1}) and Eq. (\ref{Eq:secular2}) in order to obtain the general 
eigenvalues when both the Zeeman and Rashba terms are present.  Here one needs to have additional 
specifications about the 2DEG since there are more unknown coefficients than there are secular equations.  
This makes the tilted-field case difficult to deal with. A probable additional equation would be 
the normalization condition but establishing it is not straightforward because of the coupling 
of adjacent orthogonal states. Due to these challenges, some researchers resorted to using perturbation 
theory \cite{Lipparini2006}, the continued fraction numerical method \cite{Alves2009}, and a 
quasi-classical approach \cite{Bychkov1990}. But as far as we know, an exact solution for the 
tilted-magnetic-field case has not yet been done. In this work we will take advantage of Landau 
level crossings in order to obtain an analytical solution to the problem.

When $\vec{B}$ is oriented normal to the 2DEG plane, energy level crossings are observed.  These 
crossings are caused by the widening energy gap $\Delta$ between opposite spins of the same Landau 
level $n$.  The gap widens due either to an increasing Rashba SOI strength \cite{Shen2004, Lipparini2006} 
or magnetic field \cite{Jiang2009} until opposite spins of different Landau levels intersect, that is, 
$E_{n - 1 \uparrow}$ crosses with  $E_{n \downarrow}$ while $E_{n + 1 \downarrow}$ crosses with $E_{n \uparrow}$.  
The state $|n - 1, \uparrow \rangle$ becomes closer to $|n, \downarrow \rangle$ than 
to $|n - 1, \downarrow \rangle$.  We invoke that $|n - 1, \uparrow \rangle$ and $|n, \downarrow \rangle$ 
are equally probable.  This also applies to $|n + 1, \downarrow \rangle$ and $|n, \uparrow \rangle$. 
Since the coefficients $a_n$ and $b_n$ measure the probability of obtaining the state $\phi_n$,
we hypothesize that a particular solution can be based on the following
\begin{equation}
|b_n|^2 = |a_{n-1}|^2 \qquad \longleftrightarrow \qquad ib_n = a_{n-1},
\label{Eq:ps2}
\end{equation}
\begin{equation}
|b_{n+1}|^2 = |a_n|^2 \qquad \longleftrightarrow \qquad b_{n + 1} = -i a_n.
\label{Eq:ps1}
\end{equation}
\noindent Using Eq.\ (\ref{Eq:ps1}), the secular Eq. (\ref{Eq:secular1}) becomes
\begin{equation}
\left(n + \frac{1}{2} + \Omega_z - E \right) a_n + \Upsilon \sqrt{n + 1} a_n + \Omega_- b_n = 0.
\label{Eq:secular1a}
\end{equation}
\noindent Similarly, substituting Eq.\ (\ref{Eq:ps2}), Eq.\ (\ref{Eq:secular2}) becomes
\begin{equation}
 \Upsilon \sqrt{n} b_n + \Omega_+ a_n + \left(n + \frac{1}{2} - \Omega_z - E \right) b_n = 0.
\end{equation}
\noindent Therefore, we obtain
\begin{equation}
 a_n = -\left(\frac{n + \frac{1}{2} - \Omega_z - E + \Upsilon \sqrt{n}}{\Omega_+} \right) b_n.
\label{Eq:a_s}
\end{equation}
\noindent Equation (\ref{Eq:a_s}) is valid in the tilted-magnetic-field case only. Otherwise we need 
to set $\Omega_+ = \Omega_- = 0$ in Eqs.\ (\ref{Eq:secular1}) and (\ref{Eq:secular2}).

Substituting Eq.\ (\ref{Eq:a_s}) into Eq.\ (\ref{Eq:secular1a}), we have
\begin{equation}
\hspace{-0.7cm}\left(\left(E_Z^+ \right)_z - E + \Upsilon \sqrt{n + 1} \right) \left(\left(E_Z^- \right)_z- E + \Upsilon \sqrt{n} \right) - \, \Omega_+ \Omega_- = 0,
\end{equation}
\noindent where $(E_Z^{\pm})_z = n + \frac{1}{2} \pm \Omega_z$ which is the $z$-component of Eq.~ (\ref{Eq:EZ}). This system of equations leads 
to a quadratic equation in $E$ which can be expressed as
\begin{eqnarray}
E^2 - \left\{2 \left(n + \frac{1}{2} \right) + \Upsilon \left(\sqrt{n} + \sqrt{n + 1} \right) \right\} E + \left(n + \frac{1}{2}\right)^2 \nonumber \\
+ \, \Upsilon^2 \sqrt{n} \sqrt{n + 1} + \left(n + \frac{1}{2} \right) \Upsilon \left(\sqrt{n}  + \sqrt{n + 1} \right)\nonumber\\
 - \, \Omega_z^2 + \Omega_z \Upsilon \left (\sqrt{n} - \sqrt{n + 1} \right) - \Omega_+ \Omega_- = 0, 
\label{Eq:quadratic}
\end{eqnarray}
\noindent with the following roots
\begin{eqnarray}
E = n + \frac{1}{2} + \frac{\Upsilon}{2} \left(\sqrt{n} + \sqrt{n + 1} \right) \nonumber\\
\pm \frac{1}{2} \sqrt{\left[\Upsilon \left(\sqrt{n + 1} - \sqrt{n} \right) + 2 \Omega_z \right]^2 + 4 \Omega_+ \Omega_-}.
\label{Eq:Evalues}
\end{eqnarray}
\begin{figure}[t]
\vspace{0.64cm}
\centering
\includegraphics[width = 3.1 in]{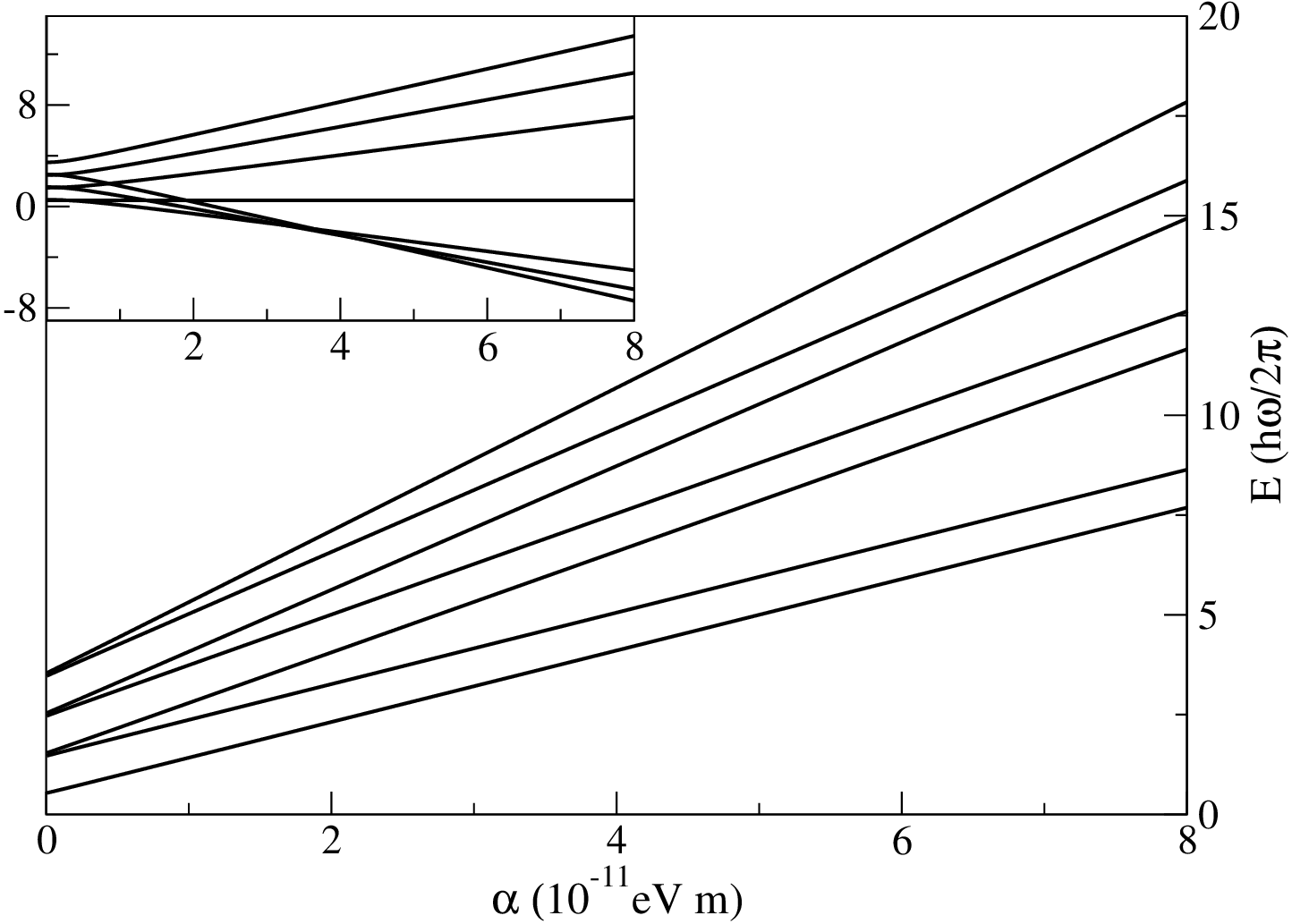}
\caption{The eigenvalues as a function of the Rashba parameter.  Here $B = 0.1$ T.  The main frame corresponds
to $\theta = 45^{\circ}$ while the inset to $\theta = 0$.}
\label{fig:Ea}
\end{figure}
\noindent The effect of tilting, based on Eq.\ (\ref{Eq:Evalues}), is to increase the gap between the spin 
up and spin down states. This results into the suppression of the crossings (See Fig.~\ref{fig:Ea}.) that are 
observed when $\vec{B}$ is perpendicular to the 2DEG plane (See inset.).  
This is true for the range of the Rashba strength studied in Fig.~\ref{fig:Ea} and for the case  when
$0 < \theta \leq 80^{\circ}$.  
In the presence of an in-plane 
component\footnote{Here ``in-plane'' and ``perpendicular'' are with respect to the 2DEG while the angle 
$\theta$ is measured in reference to the axis normal to the system's plane.} of the magnetic field 
$B_{\parallel}$, the Landau levels are now substantially separated as shown in the main frame of the 
same figure.  The same absence of crossings due to $B_{\parallel}$ was also shown in Ref.\ 
\cite{Bychkov1990} where they considered a quasi-classical limit. 
However, our preliminary investigation shows that beyond $80^{\circ}$ crossings reappear.

Inspecting Fig.\ \ref{fig:Ea} closely, for a given $n$ the splitting $\Delta$ is minimal, but not zero, 
when $\alpha = 0$.  At this point the splitting originates solely from the Zeeman term and the energies 
of the $|n, \uparrow \rangle$ and $|n, \downarrow \rangle$ are close to each other.  As the Rashba parameter 
is increased, the gap widens until we can observe that the gap between $| n+1, \downarrow \rangle$ and 
$|n, \uparrow \rangle$ becomes smaller than that of the $|n, \uparrow \rangle$ and $|n, \downarrow \rangle$.  
It is remarkable that the $| n+1, \downarrow \rangle$ and $|n, \uparrow \rangle$ indicate parallelism assuring 
that crossings are unlikely to occur even at strong Rashba SOI. 

\section{Density of States}
In analyzing the properties of a 2DEG we usually refer to the density of states (DOS) which indicate the 
number of one-electron levels in the energy range $E$ to $E + dE$ \cite{Ashcroft1976}.  In this work, we will be 
employing its Gaussian form
\begin{equation}
\mbox{DOS}(E) = \frac{eB \mbox{cos}\theta}{h} \sum_n \left(\frac{1}{2\pi} \right)^{1/2} \frac{1}{\Gamma} \, \exp \left[-\frac{(E - E_n)^2}{2 \Gamma^2} \right].
\label{Eq:DOS} 
\end{equation}
\noindent The tilt angle $\theta$ affects both the magnitude and the center $E_n$ of the Gaussian peaks. The 
parameter $\Gamma$ denotes the Landau level broadening width. Any variations of the DOS with reference to the 
different tunable parameters involved can be used to predict and explain the behavior of thermodynamic quantities.

\subsection{Rashba interaction effects}
\begin{figure}[t]
\vspace{0.64cm}
\centering
\includegraphics[width = 3.1 in]{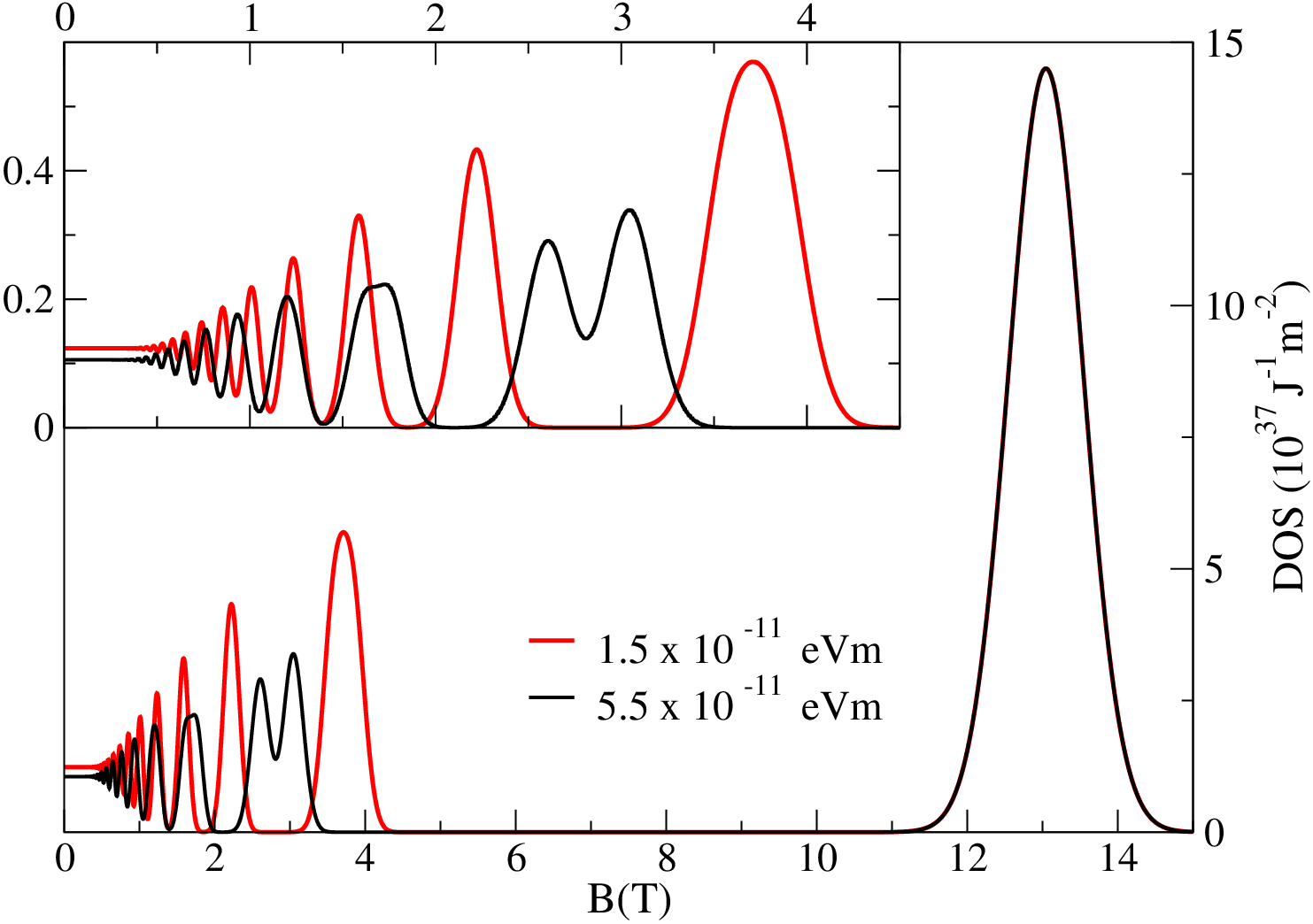}\\
\caption{The density of states as a function of $B$ for $E \approx 13.18$ meV and for two different Rashba parameters.  
Note that the two curves coincide at $B\gtrsim 4.5$ T. The plot is taken for $\Gamma = 0.5$ meV and $\theta = 22.5^{\circ}$.}
\label{fig:DBa}
\end{figure}
\begin{figure}[t]
\vspace{0.64cm}
\centering
\includegraphics[width = 3.1 in]{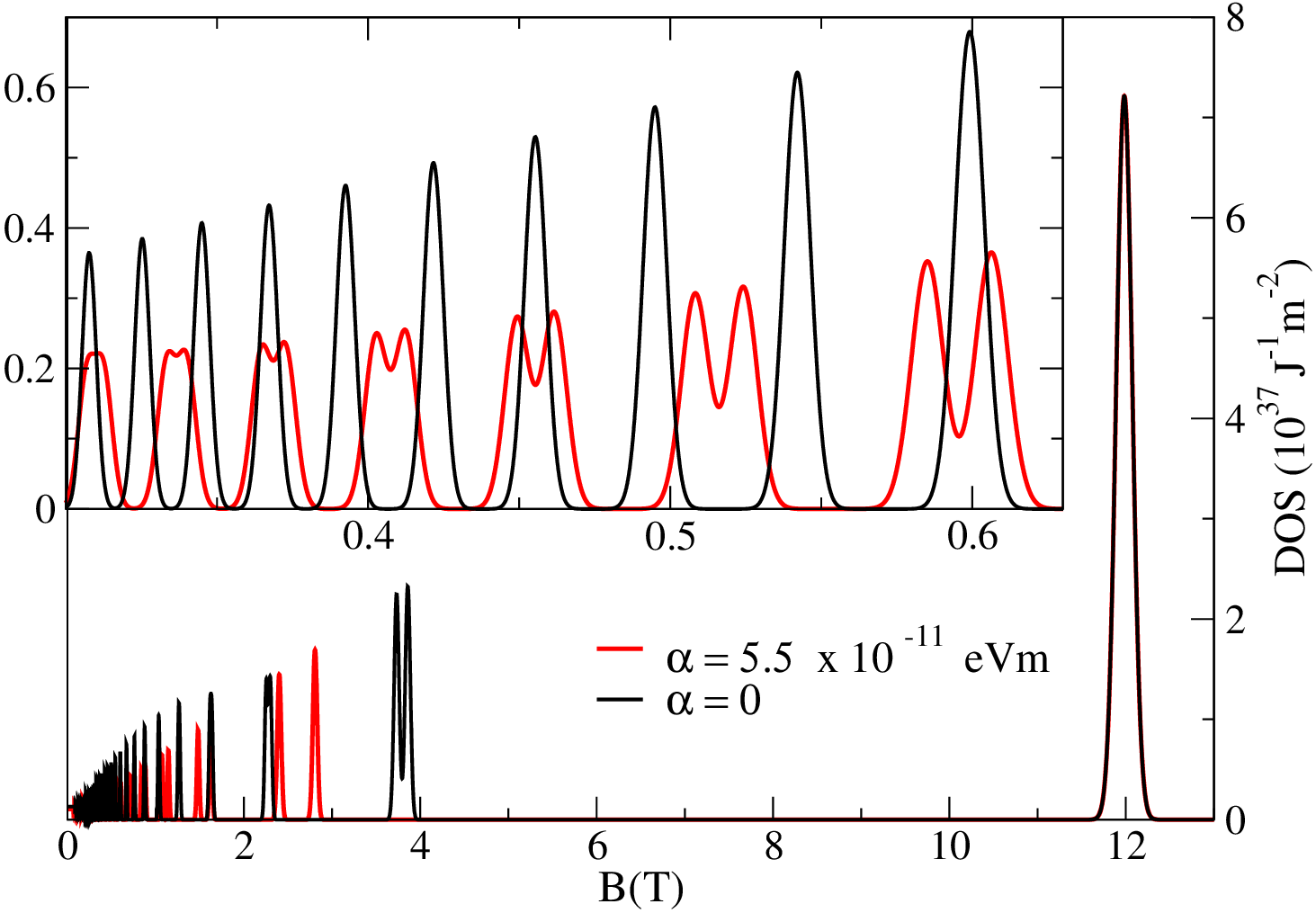}\\
\caption{The density of states as a function of $B$ for $E \approx 13.18$ meV in the absence and presence of Rashba SOI. 
Here $\theta = 80^{\circ}$ and $\Gamma = 0.1$ meV.}
\label{fig:WithNoR}
\end{figure}

For a particular energy level, we plot the DOS versus $B$ for two different Rashba SOI strengths in 
Figs.\ \ref{fig:DBa} and \ref{fig:WithNoR}.  In addition to the Zeeman term, the presence of the Rashba 
interaction usually widens the spin-splitting gap as can be seen in the eigenvalues obtained in 
Eq.\ (\ref{Eq:Evalues}). This is evident in Fig.\ \ref{fig:DBa} especially in the region 2 T $ < B < $ 4 T.  
We can attribute the splitting for different $\alpha$ to the Rashba interaction exclusively since both curves 
are under the same magnetic field region.  At these moderately strong fields, the Zeeman splitting is hardly 
noticeable. In contrast, at a very strong magnetic field region ($B > 10$ T),  changes in the Rashba 
interaction strength will have a negligible effect on the DOS.  This is expected since the Zeeman term 
is dominant at large $B$. The precise coincidence of the two curves at very strong $B$ region supports 
this. This is further emphasized in Fig.\ \ref{fig:WithNoR} where the Rashba term was completely turned 
off for the solid line.  At the main frame, the splitting at $B > 2$ T originates from the Zeeman term.  
From the inset, we notice that at weak magnetic fields ($B < 1$ T, see inset of Fig.\ \ref{fig:WithNoR}), 
the spin-splitting brought about by the Rashba SOI persists while that arising from the Zeeman 
interaction becomes negligible.

\subsection{Landau level broadening effects}
\begin{figure}[t]
\vspace{0.64cm}
\centering
\includegraphics[width = 3.1 in]{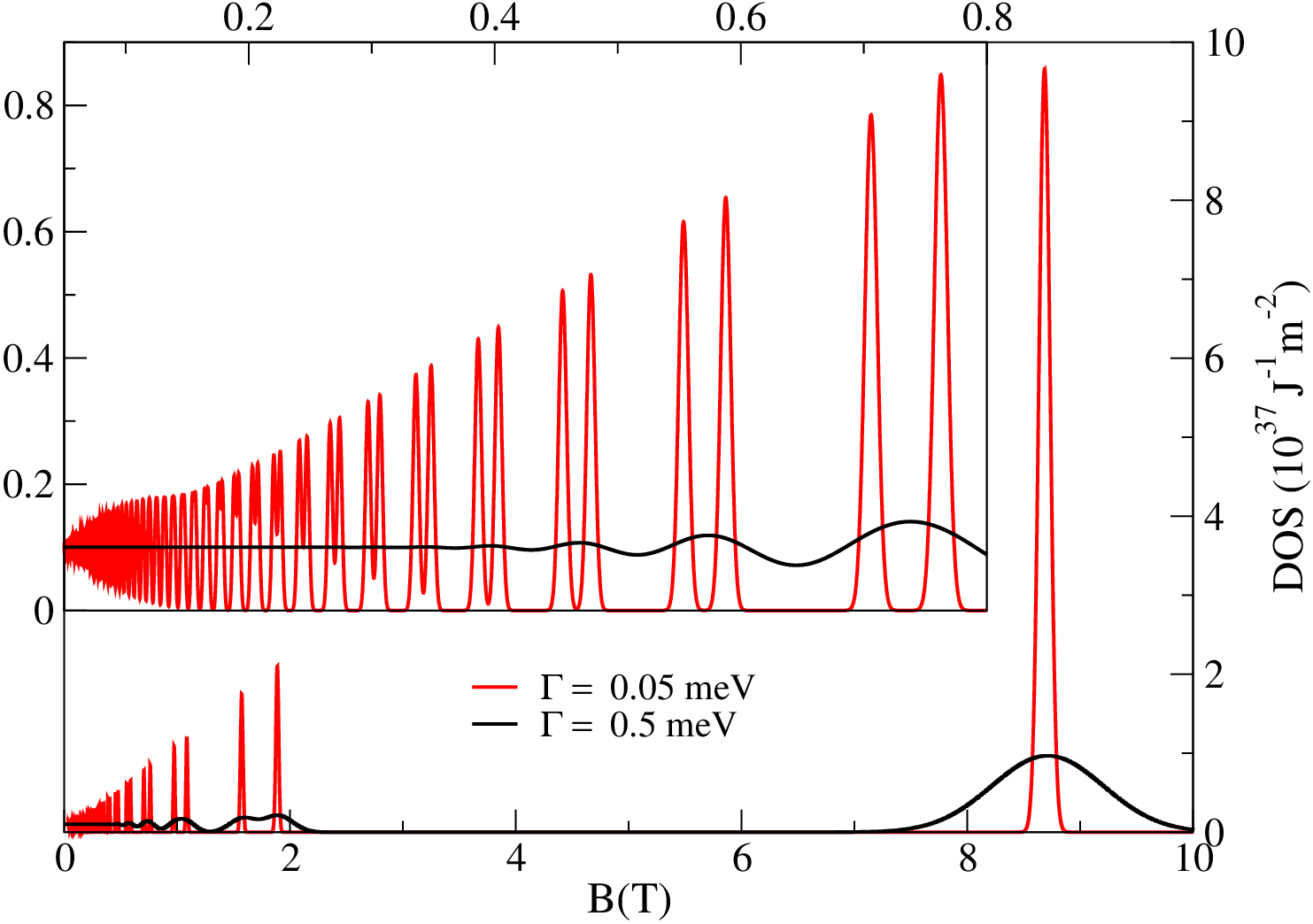}\\
\caption{The density of states as a function of $B$ for $E \approx 8.79$ meV for different 
Landau level widths.  Here $\alpha = 5.5 \times 10^{-11}$ eVm and $\theta = 22.5^{\circ}$.}
\label{fig:DBw}
\end{figure}

The broadening parameter is considered to be a measure of disorder in the system \cite{Li1989}. 
Based on Fig.\ \ref{fig:DBw} wider broadening obscures the oscillations and spin splittings 
at the lower $B$ region. Together with Fig.\ \ref{fig:DBa}, we can infer that both the 
reduction of disorder in the system (smaller $\Gamma$) and increase of the heterostructure 
asymmetry (larger $\alpha$) pave the way for the oscillations at the weak field region to manifest. 
A similar obscuring effect was observed for the spinless case \cite{GV2008}.  Hence, samples with smaller $\Gamma$
will likely be able to exhibit the pronounced splittings.

\subsection{Tilt effects}
\begin{figure}[t]
\vspace{0.64cm}
\centering
\includegraphics[width = 3.1 in]{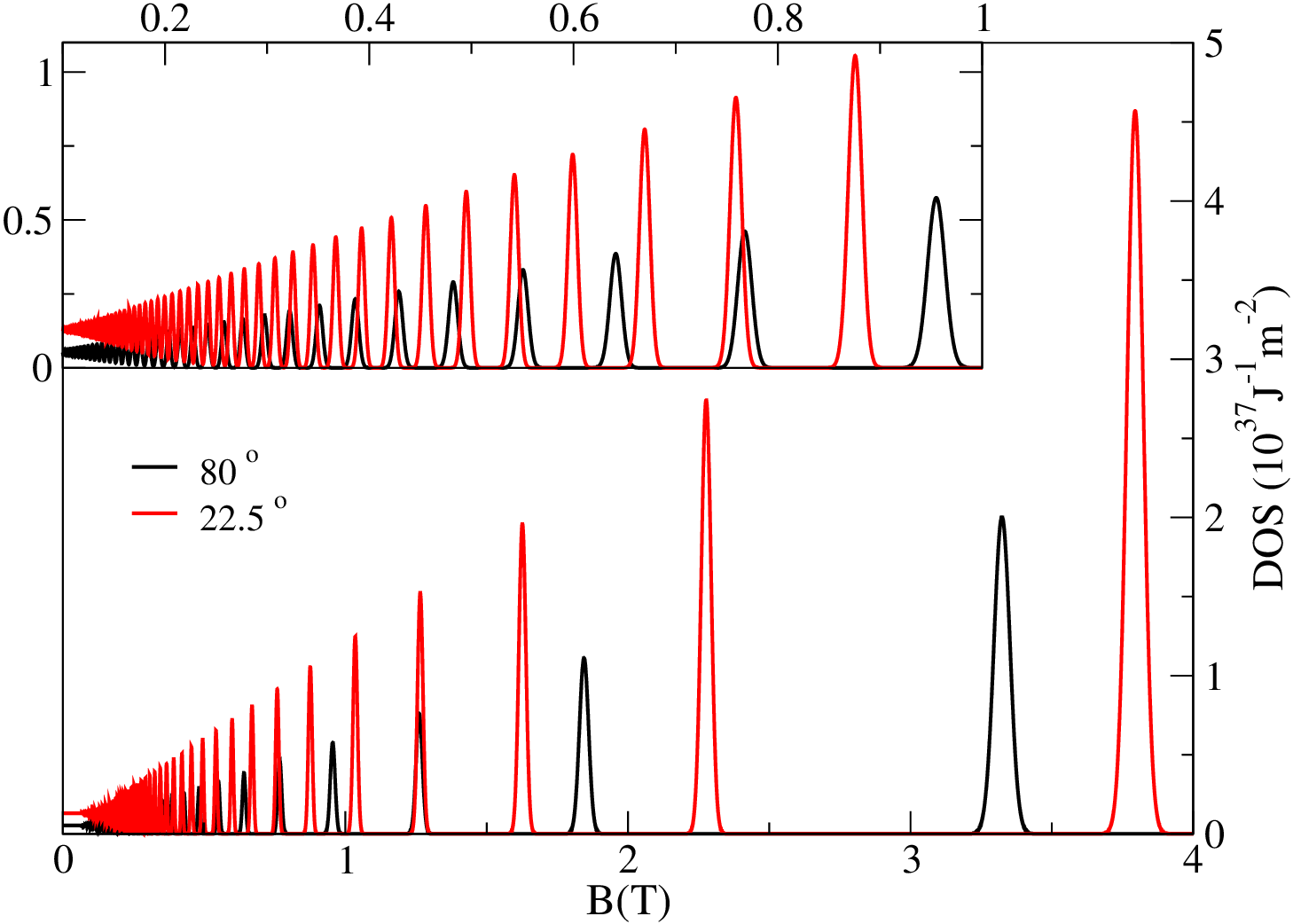}\\
\caption{The density of states as a function of $B$ for $E \approx 13.18$ meV at different tilt angles when the 
Zeeman term is neglected.  Here $\alpha = 5.5 \times 10^{-11}$ eVm and $\Gamma = 0.1$ meV.}
\label{fig:Rash_theta}
\end{figure}

\begin{figure}
\vspace{0.64cm}
\centering
\includegraphics[width = 3.1 in]{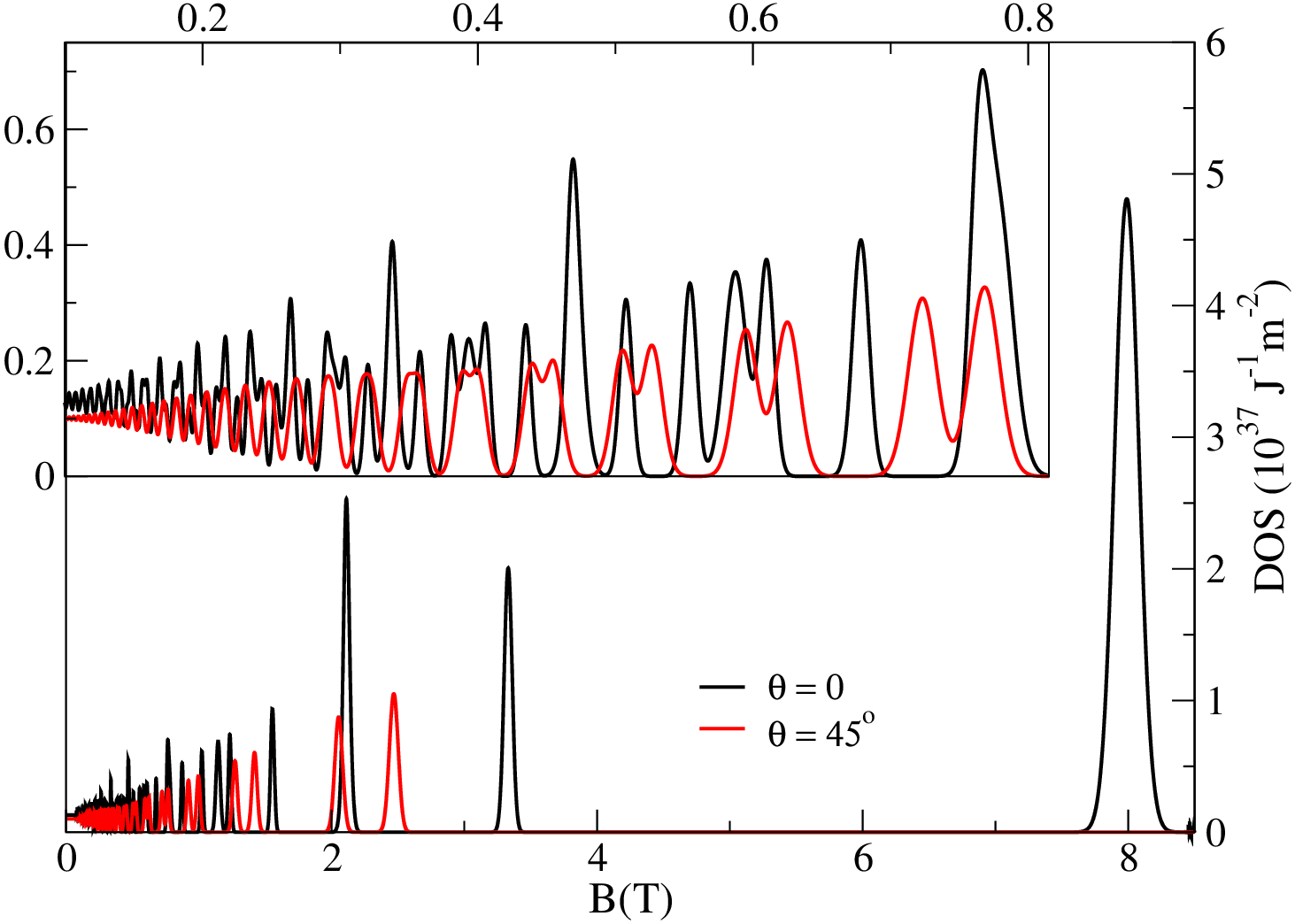}\\
\caption{The density of states as a function of $B$ for $E \approx 8.79$ meV showing the comparison of 
the no tilt scenario from the tilted case.  Here $\alpha = 5.5 \times 10^{-11}$ eVm and $\Gamma = 0.1$ meV.}
\label{fig:PerpTilt}
\end{figure}

The presence of an in-plane component of $B$ causes a non-uniform phase shift to the DOS oscillations as can be seen in Fig.\
\ref{fig:Rash_theta}. From Eq.\ (\ref{Eq:DOS}), the height of the DOS peaks is determined only by the prefactor 
$B_z = B$cos$\theta$.  This explains the taller peaks for $\theta = 22.5^{\circ}$.  The ``phase shift'' in 
the DOS oscillations that depends on the direction of the field might be related to the anisotropy measured 
in the magnetization data of Ref.\ \cite{Wilde2009}.  Since the magnetization leads to a measure of the ground state DOS, 
it is likely that the behavior of the experimental results for magnetization reflects a similar shifting of the DOS phase.

In order to compare the influence of the tilt angle, we study Fig.\ \ref{fig:PerpTilt}.  For the perpendicular case ($\theta = 0$),
the presence of taller peaks in between shorter ones is reminiscent of beating patterns observed in thermodynamic 
quantities \cite{Luo1990, Heida1998,Studenikin2005}.  The monotonic increase in the height of the DOS peaks for 
$\theta = 45^{\circ}$ can point to obscured beats.  We can deduce here that the occurence of beats can be traced 
from the crossings of energy levels which are present only in the perpendicular-magnetic-field case.  This is unlike 
the data of Ref.\ \cite{Wilde2009} where the beat structures become more prominent at large tilt angles.  We note, 
however, that the measurements reported by their group was with respect to $B_z$ only and not with respect to the 
total $B$ as is the case in this present work.  Moreover, they have fitted their empirical data with a broadening 
parameter $\Gamma \propto \sqrt{B}$ while a constant $\Gamma$ is assumed herein.  In comparison we also simulated 
the DOS using the $\sqrt{B}$-dependent broadening and found stronger oscillations when $B < 1$ T. Smeared and 
broader Gaussian peaks are observed when $B > 1$ T.  This is expected because narrower widths would be anticipated 
in weak fields and wider widths in stronger fields.  Nonetheless there were still no noticeable beating patterns.

\section{Conclusions}
A two-dimensional electron gas under a tilted magnetic field with Rashba and Zeeman interactions was studied.  
An exact solution was obtained for the case when opposite spin states of neighboring Landau levels are equally 
probable.  Eigenvalue crossings were not observed for the tilt angles studied here ($\theta \leq 80^{\circ}$).
This agrees with the numerical result of Ref.~\cite{Jiang2009} wherein the energy crossing 
degeneracies are lifted for a 2DEG without disorder.
The tilting of magnetic fields for ($0 < \theta \leq 80^{\circ}$) resulted to quenching 
of the beating patterns in the density of states.
We conclude that it is not the unequal spacing between energy levels but 
the presence of crossings that directly cause the beats observed in literature.  The persistence of the 
spin-splitting in weak and moderately strong magnetic fields brought about by the Rashba interaction is 
more perceptible in systems with less disorder.  Although the Rashba interaction causes splittings, it 
also reduces the amplitude of the oscillations similar to the effect of the Landau level broadening.  
The Zeeman contribution to the splitting is comparable to that of the Rashba term at moderate fields 
and becomes dominant at strong fields.  The determination of these different regimes as well as the 
effects of the field's tilt angle will enable the fitting of the 2DEG according to their desired state and application.
%
%
%

This work shows that tuning the Rashba interaction with tilted fields yields 2DEG 
behavior which varies differently if the magnetic field is purely perpendicular. 
While this manuscript was in preparation, the authors came across new results in 
Ref.~\cite{Xia2011} which reveal a drastic change in the behavior of the longitudinal 
resistance from the perpendicular to the case with tilted fields, even for small angles.
How this transition is approached must come from self-organizing electron interactions 
in addition to the Rashba and Zeeman interactions.

\section*{Acknowledgment}

R. Gammag is grateful to the Commission on Higher Education (CHED) for the Ph.D. scholarship provided
through the CHED - National Institute of Physics as a Center of Excellence Program.





\bibliographystyle{elsarticle-num}



\end{document}